\documentclass[12pt]{article}
\usepackage{geometry}
\usepackage{a4}
\usepackage{graphicx}
\usepackage{epsf}
\usepackage{amsmath}
\usepackage{amssymb}
\usepackage{cite}
\usepackage{multirow,tabularx}
\usepackage{appendix}
\newcommand{\be}{\begin{equation}}
\newcommand{\ee}{\end{equation}}

\newcommand{\Rmnum}[1]{\expandafter\@slowromancap\romannumeral #1@}
\newcommand{\bea}{\begin{eqnarray}}
\newcommand{\eea}{\end{eqnarray}}

\begin{document}
\def\A{{\mathbb{A}}}
\def\B{{\mathbb{B}}}
\def\C{{\mathbb{C}}}
\def\R{{\mathbb{R}}}
\def\s{{\mathbb{S}}}
\def\T{{\mathbb{T}}}
\def\Z{{\mathbb{Z}}}
\def\W{{\mathbb{W}}}
\begin{titlepage}
\title{Logarithmic Black Hole Entropy Corrections and Holographic R\'{e}nyi Entropy}
\author{}
\date{
Subhash Mahapatra \footnote{subhash.mahapatra@kuleuven.be, subhmaha@imsc.res.in}  
\vskip1.6cm
The Institute of Mathematical Sciences, \\
Chennai 600113, India \\
\& \\
KU Leuven Campus Kortrijk - KULAK, Department of Physics, \\ 
Etienne Sabbelaan 53 bus 7800, 8500 Kortrijk, Belgium}
\maketitle
\abstract{The entanglement and R\'{e}nyi entropies for spherical entangling surfaces in CFTs with gravity duals can be explicitly calculated by mapping these entropies first to 
the thermal entropy on hyperbolic space and then, using the AdS/CFT correspondence, to the Wald entropy of topological black holes. Here we extend this idea by taking into account
corrections to the Wald entropy. Using the method based on horizon symmetries and 
the asymptotic Cardy formula, we calculate corrections to the Wald entropy and find that these corrections are proportional to the logarithm of horizon area.
With the corrected black hole entropy expression, we then find corrections to the R\'{e}nyi entropies.
We calculate these corrections for both Einstein as well as Gauss-Bonnet gravity duals.
Corrections with logarithmic dependence on the area of the entangling surface naturally occur at the order $G_{D}^0$ and it seems to be a general feature of entanglement 
and R\'{e}nyi entropies for CFTs
with gravity duals. In particular, there is a logarithmic correction to the entropy in odd boundary spacetime dimensions as well.
\noindent
}
\end{titlepage}

\section{Introduction}

Quantum entanglement is one of the most remarkable properties of quantum systems which is essential to most of quantum information applications. The quantification and characterization 
of entanglement is an important problem in quantum-information science and a number of measures have been suggested for its definition in the literature \cite{Nielsen}. Two such measures are 
entanglement and R\'{e}nyi entropies. Formally, given a subsystem $v$ and its compliment $\bar{v}$, the entanglement entropy of the subsystem $v$ is defined by the von Neumann entropy
of its reduced density matrix $\rho_v$ 
\begin{eqnarray}
S_{EE}= - Tr[\rho_{v} \ln{\rho_v}]
\label{EEdefinition}
\end{eqnarray}
where $\rho_v$ is obtained by tracing out the degrees of freedom of subsystem $\bar{v}$ from the full density matrix. Similarly, R\'{e}nyi entropy is defined by a  one-parameter
generalization of the von Neumann  entropy
\begin{eqnarray}
S_{q}= \frac{1}{1-q} \ln{Tr[\rho_{v}^q]}
\label{REdefinition}
\end{eqnarray}
where $q$ is a positive real number. Given the above definition, one finds that $\lim_{q\rightarrow1} S_q=S_{EE}$, \textit{i.e}, entanglement entropy is recovered from R\'{e}nyi entropy by taking
the limit $q\rightarrow 1$.
\\
The properties of entanglement and R\'{e}nyi entropies have been the subject of intense investigations in the last two decades and form the basis for many applications ranging from 
condensed matter physics to quantum gravity. The standard way of calculating these entropies in quantum field theories is the replica method \cite{Calabrese}. In this method, one maps
the trace of the $q$-th power of the density matrix $\rho_{v}$ to the partition function on a singular $q$-folded Riemann surface. Geometrically, this $q$-fold space is a flat 
cone with angle deficit $2\pi(1-q)$ at the entangling surface and the Euclidean path integral over fields defined on this $q$-fold space can be explicitly computed for a few simple cases.
However, for generic quantum field theories the computation of the partition function on the singular surfaces is rather difficult, which limits the usefulness of the replica method.

On the other hand, recent developments in the AdS/CFT correspondence \cite{Maldacena} have suggested an elegant and geometric way of computing entanglement entropy in conformal field 
theories (CFT) which have gravity duals.
In the seminal work of Ryu and Takayanagi (RT) \cite{Ryu0603}\cite{Ryu0605}, the  entanglement entropy of a $d$-dimensional boundary CFT was conjectured to be given by the area of a minimal 
surface in the bulk AdS space as
\begin{eqnarray}
S_{EE}= \frac{Area(\gamma_v)}{4 G_D}
\label{RTEE}
\end{eqnarray}
where $\gamma_v$ is the $(d-1)$-dimensional minimal-area hypersurface, which extends into the bulk spacetime and shares the same boundary $\partial v$ of the subsystem $v$. This proposal
has been extensively tested for a variety of systems and by now there is a good amount of evidence to support this conjecture \cite{Headrick1312}. Indeed, further indications for the correctness of this
conjecture was provided in \cite{Lewkowycz1304}, where the generalized replica method was applied in the bulk AdS space to prove the RT conjecture.

An alternative approach to calculate the holographic entanglement entropy, which works only for spherical entangling surfaces, was proposed in \cite{Casini1102}. 
In \cite{Casini1102}, it was observed that the reduced density matrix for a spherical entangling region in flat space can be conformally mapped to the thermal density matrix on the 
hyperbolic space; or equivalently the original entanglement entropy can be mapped to thermal entropy on the hyperbolic space \cite{Casini1007}. Then, using the AdS/CFT correspondence,
the latter thermal entropy can be again mapped on the dual gravity side to black hole entropy of certain topological black holes with hyperbolic horizons. This observation therefore also provided an 
alternative derivation of the RT conjecture for a spherical entangling surface in $d$-dimensional CFT. Moreover in \cite{Hung1110}, it was observed that the idea of \cite{Casini1102} can be 
further extended to study the R\'{e}nyi entropy for a 
spherical entangling region. This is highly useful from the practical point of view, since no concise prescription to calculate R\'{e}nyi entropy holographically is known yet.

This procedure of mapping entanglement and R\'{e}nyi entropies of the boundary CFT to black hole entropy have many advantages. The most important advantage perhaps, apart from providing
computational simplicity, is that one can easily extend this procedure to calculate R\'{e}nyi entropy for higher dimensional CFTs. This is important since most of the work regarding
holographic R\'{e}nyi entropy has been limited to two-dimensional boundary CFTs \cite{Headrick1006}\cite{Faulkner1303}. Moreover, this mapping of R\'{e}nyi entropy can be easily 
generalized to boundary theories which are dual to higher derivative gravity theories.

An essential point which comes in the derivation of \cite{Casini1102}\cite{Hung1110} is the notion of black hole entropy, which can be computed using the standard Wald formula.
Using the Wald entropy, the above procedure indeed reproduces independent known results of R\'{e}nyi entropy. However 
an important point, which we want to stress here and that will play a significant role in this paper, is that the Wald entropy only gives the leading order contribution to black hole entropy. 
There are additional subleading quantum corrections to the usual Wald entropy, which are generally proportional to the logarithm of horizon area.
Indeed, recent activities in 
most quantum gravity models, including string theory and loop quantum gravity (LQG) or methods based on diffeomorphism symmetry arguments, have predicted logarithmic corrections 
to black hole entropy, with a general expression like
\begin{eqnarray}
S = S_{Wald} + \textbf{C}  \ln{S_{Wald}} +...
\label{correctedentropy}
\end{eqnarray}
The coefficient $\textbf{C}$ has been calculated for a variety of cases. For  example, for the Schwarzschild black hole the coefficient $\textbf{C}$ was found to be equal to $-3/2$ 
in LQG \cite{Kaul}. There, the horizon was treated as a boundary spacetime which was quantized 
within the ``quantum  geometry'' program.  Similarly in string theory, using the Euclidean gravity approach,
logarithmic corrections to the entropy of extremal and non-extremal asymptotically flat black holes have been computed in \cite{Sen1005}-\cite{Bhattacharyya1210}, also see 
\cite{Solodukhin9407}-\cite{Fursaev9412}. 
There, the logarithmic correction arises at the one loop level from the massless fields of the theory and the coefficient $\textbf{C}$ was found to be dependent on the total number of spacetime dimensions as well 
as number of massless fields. A simple comparison showed disagreement on the value of $\textbf{C}$ in these two quantum gravity models \footnote{Black hole entropy can
also acquires logarithmic corrections due to thermal fluctuations in the black hole extensive parameters.  More details on this can be found in
\cite{Das0111}-\cite{Mahapatra}.}.

The black hole entropy and its logarithmic correction can also be calculated using an approach based on the argument of diffeomorphism symmetry on the horizon. This approach, using the work 
of \cite{Brown}, was initiated in \cite{Strominger} and later emphasized in \cite{Carlip9806}-\cite{Carlip9906} to calculate black hole entropy. In this approach one first identifies
a set of vector fields (based on some physical consideration) and then construct an algebra for the Fourier modes of the charges (corresponding to an appropriate diffeomorphism symmetry) from these vector
fields. Using this algebra, which usually has a two dimensional Virasoro algebra like structure, one then extracts the central charge and the zero mode from it and then finally use the
Cardy formula \cite{Cardy1986} for the asymptotic density of states to compute black hole
entropy. It has been shown in \cite{Carlip9806}-\cite{Carlip9906} that this approach does indeed correctly reproduce the expression for the black hole entropy. Moreover in \cite{Carlip2000}, 
it was shown that this symmetry based approach can be further used to calculate the logarithmic correction to the BTZ black hole and it was found that $\textbf{C}=-3/2$ \footnote{This result was based on 
certain assumptions. We will discuss these assumptions in more details in section 4.}.

Importantly, the above procedure can also be generalized to compute logarithmic corrections to the entropy of hyperbolic black holes which are relevant for our
purpose here, especially since it is unclear (at-least to the author) how to calculate logarithmic corrections to entropy of these black holes in LQG or in string 
theory context. In this paper, using the methodology of \cite{Majhi1204}, we calculate these
corrections for two gravity theories, namely Einstein and Gauss-Bonnet gravity. We find $\textbf{C}=-3/2$ for both cases. However, we want to stress from the outset that the whole purpose 
for this exercise is to show at-least one method by which corrections to black hole entropy of hyperbolic black holes can be explicitly calculated and that it is nonzero.
Although, important in its own way,
it is not our main objective to settle down the controversial nature of $\textbf{C}$ here. Instead, we want to probe the effects of this $\textbf{C}$ on the R\'{e}nyi entropy.
For this reason, we will work with arbitrary $\textbf{C}$ in this paper for most of the time. 

Using the corrected black hole entropy expression (eq.(\ref{correctedentropy})) into the prescription of \cite{Casini1102}\cite{Hung1110}, we find that there are corrections to the standard 
results of R\'{e}nyi entropies. These corrections are both logarithmic as well as non-logarithmic in nature. For a two dimensional CFT, we find that the correction terms in the R\'{e}nyi entropy
are function of the logarithmic of the central charge as well as index $q$. However, the dependence on the size of the system is doubly logarithmic. For higher dimension CFTs, R\'{e}nyi entropy 
is a complicated function of $q$. For the entanglement entropy our results simplify tremendously. We
find that correction term in the entanglement entropy is proportional to $\textbf{C}$ times the logarithm of its standard expression. Moreover, this result is the same in all dimensions.
Interestingly, at the order $G_{D}^0$, the correction term in the entanglement entropy depends logarithmically on the area of the spherical entangling surface. This result holds for
odd boundary dimensions as well. This is important since quantum corrections to the holographic entanglement entropy are expected to be of order $G_{D}^0$. Our analysis suggests a similar kind 
of corrections in the R\'{e}nyi entropy too. With a Gauss-Bonnet black
hole as a gravity dual, the holographic R\'{e}nyi entropy is found to be a complicated function of two distinct central charges. However, the entanglement entropy depends only on one central charge.
The correction terms in the entanglement entropy is again found to be proportional to $\textbf{C}$ times the logarithm of its standard expression.

It is also well known that R\'{e}nyi entropy satisfies few a inequalities involving the derivative with respect to $q$ \cite{Beck}-\cite{Zyczkowski}. We find that for higher 
dimensional CFTs, these inequalities are again satisfied even with correction terms
provided the coefficient $\textbf{C}$ is not very large. However for two dimensional CFTs, a few of these inequalities can be violated in limit of small central charge.
Although for large central charge, which is the case for the boundary CFT with gravity dual,
these inequalities are again found to be satisfied. 

The paper is organized as follows : In the next section, we review the main ideas of \cite{Casini1102}\cite{Hung1110} to relate the R\'{e}nyi entropy to black hole entropy. In section 3, 
we highlight the necessary steps to calculate the asymptotic form of density of states using the two dimensional conformal algebra. In section 4, we first construct the Virasoro algebra 
having a central extension on the black hole horizon and then use the expression of the density of states to calculate the black hole entropy. In the process we calculate the logarithmic correction to the 
entropy of AdS-Schwarzschild and Gauss-Bonnet black hole. In section 5, we analyze the holographic R\'{e}nyi entropy in detail and discuss the nature of the correction terms.
Finally, we conclude by summarizing our main results in section 6.

\section{Holographic entanglement and R\'{e}nyi entropies}
In this section, we review some aspects of holographic entanglement and R\'{e}nyi entropies which will be the focus of this paper. As mentioned in the introduction, we follow the
prescription of \cite{Casini1102}\cite{Hung1110} to calculate these entropies for a spherical entangling region in the boundary CFT using the AdS/CFT correspondence.  Let us first briefly 
discuss the work done in these papers to set the stage. 

Consider a $d$-dimensional CFT on Minkowski space and choose a spherical entangling surface of radius $R$ as a subsystem. The computation of entanglement entropy of this subsystem with the
rest of the system can be performed by calculating the reduced density matrix $\rho_v$. However the authors in \cite{Casini1102}, using the conformal structure of the theory, mapped this problem
of entanglement entropy to the thermal entropy on a hyperbolic space $R\times H^{d-1}$. They showed that the causal development of the ball enclosed by the
spherical entangling surface can be mapped to a hyperbolic space $R\times H^{d-1}$; with curvature of $H^{d-1}$ space given by the radius of the spherical entangling region $R$. 
An important point of this mapping was that the vacuum of the original CFT mapped to a thermal bath with temperature 
\begin{eqnarray}
T_0=\frac{1}{2\pi R}
\label{T0temp}
\end{eqnarray}
on the hyperbolic space. Now relating the density matrix $\rho_{therm}$ in the new spacetime $R\times H^{d-1}$ to the old spacetime $\rho_v$ by the unitary transformation:
$\rho_v=U^{-1}\rho_{therm}U$, we get
\begin{eqnarray}
\rho_v=U^{-1} \frac{e^{[-H/T_0]}}{Z(T_0)}U
\label{rhov}
\end{eqnarray}
where ${Z(T_0)}=Tr[e^{-H/T_0}]$. For the R\'{e}nyi entropy we also need the $q$'th power of $\rho_v$. From the above equation, we get
\begin{eqnarray}
\rho_{v}^q=U^{-1} \frac{e^{[-qH/T_0]}}{Z(T_0)^q}U
\label{rhovq}
\end{eqnarray}
Taking the trace of both sides of eq.(\ref{rhovq}), we get
\begin{eqnarray}
Tr[\rho_{v}^q]=\frac{Z(T_0/q)}{Z(T_0)^q}
\label{tracerhovq}
\end{eqnarray}
as $U$ and its inverse cancels each other upon taking trace. Now, using the definition of R\'{e}nyi entropy as in eq.(\ref{REdefinition}), we arrive at
\begin{eqnarray}
S_{q}= \frac{1}{1-q} \bigl[\ln{Z(T_0/q)}-q \ln{Z(T_0)}  \bigr]
\end{eqnarray}
The above expression for the R\'{e}nyi entropy can also be written in terms of the free energy $F(T)=-T\ln{Z(T)}$:
\begin{eqnarray}
S_{q}= \frac{q}{1-q}\frac{1}{T_0} \bigl[F(T_0)-F(T_0/q)  \bigr]
\label{Sqfreeenergy}
\end{eqnarray}
and further, using the thermodynamic relation $S_{therm}=-\partial F/\partial T$, we can rewrite the above expression as
\begin{eqnarray}
S_{q}= \frac{q}{q-1}\frac{1}{T_0} \int_{T_0/q}^{T_0} dT \  S_{therm}(T)
\label{Sqmain}
\end{eqnarray}
here, just to clarify again, $S_{therm}$ is the thermal entropy of a $d$-dimensional CFT on $R\times H^{d-1}$ while $S_q$ is the desired R\'{e}nyi entropy. Eq.(\ref{Sqmain}) was the main result
of \cite{Casini1102}\cite{Hung1110}, which relates R\'{e}nyi (and hence entanglement) entropy of a spherical entangling region in  $d$-dimensional CFT to the thermal entropy on a hyperbolic space. 
As pointed out in \cite{Casini1102}, the above analysis just mapped one difficult problem to another equally difficult problem and is not particularly useful for practical purposes. However, its true usefulness can 
be realized via the AdS/CFT correspondence. In the AdS/CFT correspondence, the thermal state of the boundary CFT corresponds to an appropriate non-extremal black hole in the bulk AdS spacetime, with
thermal entropy corresponding to black hole entropy. Therefore, using the AdS/CFT correspondence, we can relate $S_{therm}$ appearing in eq.(\ref{Sqmain}) to that of black hole entropy, 
which is relatively easy to compute. Since on the boundary side our CFT is on $R\times H^{d-1}$, its dual gravity theory will be described by a topological black hole with hyperbolic
event horizon. In any event, in this AdS/CFT approach, the R\'{e}nyi entropy is now given by the horizon entropy of the corresponding hyperbolic black hole, which can be easily
computed using Ward's standard formula. However, since we are interested in calculating the effects of corrections of black hole entropy on the
R\'{e}nyi entropy, here we will use an approach based on symmetry arguments on the horizon to calculate the black hole entropy, instead of Wald's formula. This is the topic of discussion
of the next section.

\section{Logarithmic corrections to black hole entropy from the Cardy formula}
In this section, we will describe the necessary steps to calculate the asymptotic form of density of states from a two dimensional conformal algebra. This form of density of states will be used 
in a later section to calculate the black hole entropy and, further, to compute logarithmic corrections. In this section, we will mostly follow the notations used in
\cite{Carlip2000} and refer the readers to \cite{Carlip2000} for a detailed discussion. 

We start with a standard Virasoro algebra of the two conformal field theory with central charges $c$, $\bar{c}$:
\begin{eqnarray}
\bigl[L_m,L_n \bigr] &=& (m-n)L_{m+n}+\frac{c}{12} m(m^2-1)\delta_{m+n,0}  \nonumber\\
\bigl[\bar{L}_m,\bar{L}_n \bigr] &=& (m-n)\bar{L}_{m+n}+\frac{\bar{c}}{12} m(m^2-1)\delta_{m+n,0}\  \nonumber\\
\bigl[L_m,\bar{L}_n \bigr] &=& 0
\label{Virasoro}
\end{eqnarray}
here $L_m$ and $\bar{L}_m$ are the generators of holomorphic and antiholomorphic diffeomorphisms. If $\rho(\Delta,\bar{\Delta})$ denotes the degeneracy of states carrying $L_0=\Delta$ and
$\bar{L}_0=\bar{\Delta}$ eigenvalues, then one can define the partition function on the two-torus of modulus $\tau=\tau_1 +i \tau_2$ as
\begin{eqnarray}
Z(\tau,\bar{\tau})=Tr(e^{2\pi i \tau L_0}e^{-2\pi i \bar{\tau} \bar{L}_0})=\sum_{\Delta,\bar\Delta}\biggl(\rho(\Delta,\bar{\Delta})e^{2\pi i \tau \Delta}
e^{-2\pi i \bar{\tau} \bar{\Delta}}\biggr)
\label{partitionZ}
\end{eqnarray}
Now using $q=e^{2\pi i \tau }$, $\bar{q}=e^{-2\pi i \bar{\tau} }$ and inverting the above equation, we get
\begin{eqnarray}
\rho(\Delta,\bar{\Delta}) = \frac{1}{(2\pi i)^2} \int \frac{1}{q^{\Delta+1}}\frac{1}{\bar{q}^{\bar{\Delta}+1}}Z(q,\bar{q}) dq d\bar{q}
\label{DOS}
\end{eqnarray}
where the integrals are along contours that enclose $q=0$ and $\bar{q}=0$. Therefore, if we know the partition function $Z(q,\bar{q})$, then we can use eq.(\ref{DOS}) to
determine the density of states. Now modular invariance of the theory implies that
\begin{eqnarray}
Z_{0}(\tau,\bar{\tau})=Tr\biggl[e^{2\pi i \tau (L_0-\frac{c}{24})}e^{-2\pi i \bar{\tau} (\bar{L}_0-\frac{\bar{c}}{24})}  \biggr]
\label{diffeo}
\end{eqnarray}
$Z_{0}$ is invariant under large $\tau\longrightarrow -1/\tau$ diffeomorphism. Cardy has shown that the invariance in eq.(\ref{diffeo}) is based on the general properties of two 
dimensional conformal field theory and therefore
expected to be universal. From eq.(\ref{partitionZ}) and (\ref{diffeo}), we note that 
$$Z(\tau,\bar{\tau})=e^{\frac{2\pi i c \tau}{24}}e^{-\frac{2 \pi i \bar{c} \bar{\tau}}{24}} Z_{0}(\tau,\bar{\tau})$$
using the modular invariance of $Z_0$, we get
\begin{eqnarray}
Z(\tau,\bar{\tau})=e^{\frac{2\pi i c}{24}(\tau+\frac{1}{\tau})}e^{-\frac{2\pi i \bar{c}}{24}(\bar{\tau}+\frac{1}{\bar{\tau}})} Z\biggl(-\frac{1}{\tau},-\frac{1}{\bar{\tau}} \biggr)
\label{modularinv}
\end{eqnarray}
substituting eq.(\ref{modularinv}) into eq.(\ref{DOS}), we obtain
\begin{eqnarray}
\rho(\Delta,\bar{\Delta}) = \int d\tau d\bar{\tau} \  e^{-2 \pi i \Delta \tau + \frac{2\pi i c}{24}(\tau+\frac{1}{\tau})}e^{2 \pi i \bar{\Delta} \bar{\tau}-\frac{2\pi i \bar{c}}{24}(\bar{\tau}+\frac{1}{\bar{\tau}})} 
Z\biggl(-\frac{1}{\tau},-\frac{1}{\bar{\tau}} \biggr)
\label{DOS1}
\end{eqnarray}
The above integral has the form
\begin{eqnarray}
I[a,b]=\int d\tau e^{2\pi i a\tau + \frac{2\pi i b}{\tau}} \mathcal{F(\tau)}
\end{eqnarray}
which can be evaluated by saddle point approximation. For this, we need to assume that $\mathcal{F(\tau)}$ is slowly varying near the extremum of the phase. As shown in 
\cite{Carlip2000}, this is indeed the case if one considers the situation where the imaginary part of $\tau$ is large.
Now the saddle point, obtained by extremizing the exponent on the right hand side of the above equation, is at $\tau_{0}=\sqrt{b/a}$ $\approx$ $i\sqrt{c/24\Delta}$, where we have 
assumed large $\Delta$. Expanding around this saddle point, we find
\begin{eqnarray}
I[a,b]\approx \int d\tau e^{4\pi i\sqrt{ab} + \frac{2 \pi i b}{\tau_{0}^3}(\tau-\tau_0)^2} \mathcal{F}(\tau_0) = \biggl(-\frac{b}{4a^3} \biggr)^{1/4} e^{4 \pi i \sqrt{ab}}\mathcal{F}(\tau_0)
\end{eqnarray}
and an analogous expression exists for the $\bar{\tau}$ integral. Finally, one obtains the expression for the density of states as
\begin{eqnarray}
\rho(\Delta,\bar{\Delta})\approx \biggl(\frac{c}{96\Delta^3} \biggr)^{\frac{1}{4}} \biggl(\frac{\bar{c}}{96\bar{\Delta}^3} \biggr)^{\frac{1}{4}} e^{2\pi\sqrt{\frac{c\Delta}{6}}} \ 
e^{2\pi\sqrt{\frac{\bar{c}\bar{\Delta}}{6}}}
\label{DOS1}
\end{eqnarray}
Since, for most cases of our interest, we have only one Virasoro algebra instead of two (see the next section for details), the relevant expression for density of states
is
\begin{eqnarray}
\rho(\Delta)\approx \biggl(\frac{c}{96\Delta^3} \biggr)^{\frac{1}{4}} e^{2\pi\sqrt{\frac{c\Delta}{6}}}
\label{DOS2}
\end{eqnarray}
In the above expression, the exponential term is the standard Cardy formula.  However an important
part, which will play a significant role in our analysis later on, is the term that has a power law behavior. In the next section, we will use the logarithm
of $\rho$ to calculate the entropy of the black holes. As one can anticipate, the exponential part of eq.(\ref{DOS2}) will give the usual Wald entropy and, on the other hand, the
power term will provide a logarithmic correction to the black hole entropy.

\section{Black hole entropy and Virasoro algebra from the surface term of the gravitational action}
There has been a lot of activity in understanding the black hole entropy using the symmetry based horizon CFT approach. This approach essentially assumes that the symmetries of a black hole 
horizon are enough to compute the density of states (hence the black hole entropy) at a given energy. Many avatars of this approach have been appeared in the literature, and all of them
have successfully predicted the black hole entropy expression \footnote{A certain level of arbitrariness is present in the procedure of all symmetry based approaches in order to produce
correct black hole entropy expression. We will discuss more about these arbitrariness in the following.}. A complete list of references for the later  development can be found in \cite{Majhi1111}.

Applying a similar line of symmetry reasoning, a new approach was recently proposed in \cite{Majhi1204} which is straightforward and conceptually more clear. Importantly, it does not require any ad hoc
prescription such as shifting the zero-mode energy and is also easy to implement. In this approach, the Noether currents associated with the diffeomorphism invariance of the Gibbons-Hawking boundary action, instead of the bulk gravity action, are used to construct to the Virasoro algebra at the horizon. Here, the diffeomorphisms are chosen in such a way that they leave the near-horizon structure of the metric invariant. The Virasoro algebra constructed in this way is again found to have a central extension, which upon using the Cardy formula correctly reproduces the black hole entropy expression. In this work, 
we will follow this boundary Noether current procedure of \cite{Majhi1204} to calculate the black hole entropy.
We will first discuss its general formalism and then apply this formalism to calculate the black hole entropy of two gravity theories, namely Einstein 
and Gauss-Bonnet gravity.
\\
\\
We start with the Gibbons-Hawking surface term $I_{GH}$:
\begin{eqnarray}
I_{GH} =\frac{1}{8 \pi G_D} \int_{\partial\mathcal{M}} d^{D-1}x \sqrt{-\gamma} \mathfrak{L}=\frac{1}{8 \pi G_D} \int_{\mathcal{M}} d^{D}x \sqrt{-g} \nabla_{a}(n^{a}\mathfrak{L})
\label{SGH}
\end{eqnarray}
where $\partial\mathcal{M}$ is the boundary of the manifold $\mathcal{M}$, $\mathfrak{L}$ is related to the trace of the extrinsic curvature, $n^a$ is the unit 
normal vector to the boundary, $g_{\mu\nu}$ denotes the bulk metric and $\gamma_{\mu\nu}$ is the induced metric on the boundary. The expression for $\mathfrak{L}$ depends on the gravity 
theory under consideration. For example, in Einstein gravity it is equal to trace of the extrinsic curvature $\mathcal{K}=-\nabla_{\mu}n^\mu$. The  conserved current $J^\mu$ associated with the
diffeomorphism invariance $x^\mu \longrightarrow x^\mu + \xi^\mu$ of $I_{GH}$ can be obtained by considering the variation 
of the both sides of eq.(\ref{SGH}) as the Lie derivative. After a bit of algebra, one gets
\begin{eqnarray}
J^\mu[\xi] = \nabla_{\nu} J^{\mu\nu}[\xi]=\frac{1}{8 \pi G_D}\nabla_{\nu}\biggl(\mathfrak{L}\xi^\mu n^\nu -\mathfrak{L}\xi^\nu n^\mu \biggr)
\label{currentJ}
\end{eqnarray}
The charge corresponding to this conserved current is defined as
\begin{eqnarray}
Q[\xi] = \int_{\partial\mathcal{\varSigma}}  d\Sigma_{\mu\nu} \bigl(\sqrt{-h} J^{\mu\nu}  \bigr)
\label{chargeQ}
\end{eqnarray}
where $J^{\mu\nu}$ is called the Noether potential and $d\Sigma_{\mu\nu}=-d^{D-2}x(n_\mu m_\nu-m_\mu n_\nu)$ is the surface
element of the $(D-2)$-dimensional surface ${\partial\mathcal{\varSigma}}$ with metric $h_{\mu\nu}$. In order to discuss black hole physics, we will chose the surface ${\partial\mathcal{\varSigma}}$ 
to be near the horizon. The unit vectors $n^\mu$ and $m^\mu$ are chosen to be spacelike and timelike, respectively. 

Following \cite{Majhi1204}, we define the Lie bracket of the charges as
\begin{eqnarray}
\bigl[Q[\xi_1],Q[\xi_2]\bigr]\equiv \biggl(\delta_{\xi_1}Q[\xi_2] - \delta_{\xi_2}Q[\xi_1] \biggr)= \int_{\partial\mathcal{\varSigma}}  d\Sigma_{\mu\nu} \sqrt{-h}
\bigl(\xi_{2}^{\mu} J^{\nu}[\xi_{1}] -\xi_{1}^{\mu} J^{\nu}[\xi_{2}]  \bigr)
\label{Qalgebra}
\end{eqnarray}
Form the above equation, it is clear that we only need to know the vector field $\xi^\mu$ to determine the charge algebra. This can be done, as explained below, by choosing a appropriate
diffeomorphism which leaves the horizon structure  invariant. However in order to proceed further, let us first write the $D(=d+1)$ dimensional form of the metric as
\begin{eqnarray}
ds^2=-f(r) N^2 dt^2 + \frac{dr^2}{f(r)} + r^2 d\Omega_{ij}(x)dx^idx^j
\label{metric}
\end{eqnarray}
where $d\Omega_{ij}(x)dx^idx^j$ is the line element of the $(d-1)$-dimensional space. For our discussion in the next section, we will take this line element to be hyperbolic
but as of now this metric is
completely general. The constant $N^2$ is included in $g_{tt}$ to allow us to adjust the normalization of the time coordinate. This will be useful later on, but will not play any
significant role in the black hole entropy calculation here. To study the near horizon 
structure, defined by $f(r_h)=0$, it is convenient to choose a coordinate $r=\tilde{r}+r_h$, in which case the above metric reduces to
\begin{eqnarray}
ds^2=-f(\tilde{r}+r_h) N^2 dt^2 + \frac{d\tilde{r}^2}{f(\tilde{r}+r_h)} + (\tilde{r}+r_h)^2 d\Omega_{ij}(x)dx^idx^j
\label{metric1}
\end{eqnarray}
In the near horizon region \textit{i.e} $\tilde{r}\rightarrow0$ limit, the function $f(\tilde{r}+r_h)$ can be expanded as 
$$f(\tilde{r}+r_h)=\tilde{r} f'(r_h)+1/2\tilde{r}^2f''(r_h)+...$$
Defining the surface gravity $\kappa$ as $\kappa=N f'(r_h)/2$ and dropping the terms beyond first order in $f(\tilde{r}+r_h)$, one notice that the $(t-\tilde{r})$ part of the metric in 
eq.(\ref{metric1}) reduces to the standard Rindler metric $$ds^{2}_{\tilde{t}-\tilde{r}}=-2 \tilde{r}\frac{\kappa}{N} d\tilde{t}^2 +\frac{1}{2\tilde{r}} \frac{N}{\kappa}d\tilde{r}^2$$
where $\tilde{t}=Nt$. The unit normal vectors can be chosen as
\begin{equation}
n^\mu=(0,\sqrt{f(\tilde{r}+r_h)},0,...,0) , \ \ \ \  m^\mu=(1/N\sqrt{f(\tilde{r}+r_h)},0,0,...,0) 
\label{unitnormal}
\end{equation}
Further, in order to find $\xi^\mu$, it is more useful to first transform to Bondi like coordinates by the transformation
\begin{equation}
du=d\tilde{t}-\frac{d\tilde{r}}{f(\tilde{r}+r_h)}
\end{equation}
under which the metric in eq.(\ref{metric1}) reduces to
\begin{equation}
ds^2=-f(\tilde{r}+r_h)du^2-2dud\tilde{r} + (\tilde{r}+r_h)^2 d\Omega_{ij}(x)dx^idx^j
\label{Bondi}
\end{equation}
Now we choose our vector field $\xi^\mu$ by imposing the condition that the horizon structure remain invariant under diffeomorphism \textit{i.e} the metric coefficients
$g_{\tilde{r} \tilde{r}}$ and $g_{u \tilde{r}}$ remain unchanged. This implies the following Killing equations \footnote{It is easy to check that the other condition
$\mathcal{L}_\xi g_{u u}=0$ is trivially satisfied near the horizon.},
\begin{eqnarray}
&& \mathcal{L}_\xi g_{\tilde{r} \tilde{r}}=-2\partial_{\tilde{r}}\xi^u=0 \nonumber\\
&& \mathcal{L}_\xi g_{u \tilde{r}}=-f(\tilde{r}+r_h)\partial_{\tilde{r}}\xi^u-\partial_{u}\xi^{u}-\partial_{\tilde{r}}\xi^{\tilde{r}}=0
\end{eqnarray}
Solving the above two equations, we get
\begin{eqnarray}
\xi^u=\mathcal{S}(u,\vec{x}), \ \ \ \ \   \xi^{\tilde{r}} = -\tilde{r} \partial_{u} \mathcal{S}(u,\vec{x})
\label{xieqn}
\end{eqnarray}
where $\mathcal{S}$ is an arbitrary function and $\vec{x}$ are the coordinates on the remaining $(d-1)$ dimensional space. Now converting back to the $(t,\tilde{r})$ coordinates,
these vector fields take the form
\begin{eqnarray}
\xi^t=T-\frac{\tilde{r}}{N f(\tilde{r}+r_h)}\partial_{t}T, \ \ \ \ \  \xi^{\tilde{r}} = -\frac{\tilde{r}}{N}\partial_{t} T
\label{killingvector}
\end{eqnarray}
where $T(t,\tilde{r},\vec{x})\equiv\mathcal{S}(u,\vec{x})$. Therefore, we can evaluate the expressions for $\xi^\mu$, $Q[\xi^\mu]$ and the charge algebra $\bigl[Q[\xi_1],Q[\xi_2]\bigr]$ once the
function $T$ is given. Now, we expand this function $T$ in terms of a set of basis functions $T_m$, as
\begin{eqnarray}
T=\sum_{m} A_{m} T_{m}, \ \ \ \ A_{m}^{*}=A_{-m}
\label{Texpension}
\end{eqnarray}
As it is standard in the literature, we choose $T_m$ such that the resulting $\xi^{\mu}_{m}$ obeys the algebra isomorphic to Diff $S^1$ \textit{i.e}
\begin{eqnarray}
i \{\xi_m,\xi_n \}^\mu=(m-n)\xi^{\mu}_{m+n}
\end{eqnarray}
where $\{,\}$ is the Lie bracket. A particular choice is
\begin{eqnarray}
T_{m}=\frac{1}{\alpha} e^{i m (\alpha t + g(\tilde{r}) + \vec{p}.\vec{x})}
\label{Tm}
\end{eqnarray}
where $\alpha$ is an arbitrary parameter, $g(\tilde{r})$ is a function which is regular at the horizon and $p_{i}$ are integers. We now have all the ingredients at our disposal to calculate the 
leading Wald as well as the subleading logarithmic correction to the black hole entropy. From here on
we will concentrate on horizons with hyperbolic topology, as this will be required in the computation of holographic R\'{e}nyi entropy in the later section.

\subsection{Entropy of AdS-Schwarzschild Black hole}
We first apply the above developed formalism to a $(d+1)$-dimensional AdS-Schwarzschild black hole in Einstein gravity. As mentioned earlier, for Einstein gravity the Gibbons-Hawking 
surface term is standard and is given by the expression $\mathfrak{L}=\mathcal{K}=-\nabla_{\mu}n^\mu$. The metric is given as
\begin{eqnarray}
&& ds^2 = -f(r)N^2 dt^2+\frac{1}{f(r)}dr^2+r^2d\Sigma^{2}_{d-1} \ , \nonumber\\
&& f(r)=-1-\frac{m}{r^{d-2}}+\frac{r^2}{L^2}
\label{AdSSch}
\end{eqnarray}
where
\begin{eqnarray}
d\Sigma^{2}_{d-1}= d\theta^2 + \sinh^2\theta \ d\varOmega_{d-2}^2
\label{hyperboliclineelement}
\end{eqnarray}
is the metric on $(d-1)$-dimensional hyperbolic space with $d\varOmega_{d-2}^2$ being the line element on a unit $(d-2)$-sphere. The exact value of constant $N^2$ is not required here 
and will be specified in the next section. Substituting eq.(\ref{AdSSch}) into eqs.(\ref{killingvector}),
(\ref{currentJ}), (\ref{chargeQ}) and (\ref{Qalgebra}) and taking the near horizon limit $\tilde{r}\rightarrow0$, we get
\begin{eqnarray}
Q[\xi_m]=\frac{1}{8 \pi G_{D}} \int_{\mathcal{H}} d^{d-1}x \sqrt{h} \bigg(\kappa T_m - \frac{1}{2}\partial_{t}T_m \biggr)
\label{chargeSAdS}
\end{eqnarray}
Similarly, the algebra of the charges corresponding to $T$=$T_m$, is given by
\begin{eqnarray}
\bigl[Q[\xi_m],Q[\xi_n]\bigr]=\frac{1}{8 \pi G_{D}} \int_{\mathcal{H}} d^{d-1}x \sqrt{h}\biggl[\kappa \bigg(T_m\partial_{t}T_n-T_n\partial_{t}T_m \biggr) \nonumber\\
-\frac{1}{2}\biggl(T_{m}\partial_{t}^{2} T_{n} - T_{n}\partial_{t}^{2}T_{m} \biggr) +\frac{1}{4\kappa}\biggl(\partial_{t}T_m\partial_{t}^{2}T_n-\partial_{t}T_n\partial_{t}^{2}T_m \biggr) \biggr]
\label{chargeBraSAdS}
\end{eqnarray}
Now substituting eqs.(\ref{Tm}) and (\ref{AdSSch}) into eqs.(\ref{chargeSAdS}) and (\ref{chargeBraSAdS}), we get the final expressions as
\begin{eqnarray}
&& Q[\xi_m]=\frac{A_{d-1}}{8 \pi G_{D}}\frac{\kappa}{\alpha}\delta_{m,0} \nonumber\\
&& \bigl[Q[\xi_m],Q[\xi_n]\bigr]=-i(m-n)\frac{A_{d-1}}{8 \pi G_{D}}\frac{\kappa}{\alpha}\delta_{m+n,0} -im^3\frac{A_{d-1}}{16 \pi G_{D}}\frac{\alpha}{\kappa}\delta_{m+n,0}
\label{Q0finalSAdS}
\end{eqnarray}
where $A_{d-1}$ is the area of $(d-1)$-dimensional hyperbolic space (the horizon area). Strikingly, the above charge algebra is quite similar to the two dimensional Virasoro algebra discussed 
in the previous section. From above expressions, we obtain the central term in the algebra as
\begin{eqnarray}
&& K\bigl[\xi_m,\xi_n\bigr]=\bigl[Q[\xi_m],Q[\xi_n]\bigr]+i(m-n)Q[\xi_{m+n}] \nonumber\\
&&  \hspace{2cm} = -im^3\frac{A_{d-1}}{16 \pi G_{D}}\frac{\alpha}{\kappa}\delta_{m+n,0}
\label{centralKSAdS}
\end{eqnarray}
from which we can read off the zero mode $Q_0$ and central charge $c$ as
\begin{eqnarray}
&& Q_0 = Q[\xi_0]= \frac{A_{d-1}}{8 \pi G_{D}}\frac{\kappa}{\alpha}, \hspace{1cm} \frac{c}{12}= \frac{A_{d-1}}{16 \pi G_{D}}\frac{\alpha}{\kappa}
\label{QoCSAdS}
\end{eqnarray}
Substituting above expressions into eq.(\ref{DOS2}) with $Q_0=\Delta$ and taking log on both side, we obtain the black hole entropy
\begin{eqnarray}
&& S=\ln{\rho}\approx 2\pi\sqrt{\frac{c\Delta}{6}} + \frac{1}{4}\ln{\frac{c}{\Delta^3}} +... \nonumber\\
&& S=\frac{A_{d-1}}{4 G_D} -\frac{3}{2} \ln{\frac{A_{d-1}}{4G_D}} +... = S_{Wald}-\frac{3}{2} \ln{S_{Wald}} +...
\label{SAdSentropy}
\end{eqnarray}
We see that the leading term matches exactly with the usual black hole entropy expression. We also find a logarithmic correction to it. Interestingly, the coefficient $-3/2$ of the
logarithmic correction for the hyperbolic horizon is the same as was found for the BTZ black hole in \cite{Carlip2000}.
In order to find this coefficient we have chosen the arbitrary parameter $\alpha$ to be such that the central charge $c$ is a universal 
constant \textit{i.e} independent of the area $A_{d-1}$, as in \cite{Carlip2000}. We will say more about this condition and the coefficient of this logarithmic term at the end of this section.

\subsection{Entropy of Gauss-Bonnet black hole}
Now we move on to discuss the entropy of Gauss-Bonnet black holes. The surface term of the gravitational action can be found in \cite{Cai}\cite{Davis} and is given as 
\begin{eqnarray}
\mathfrak{L}=-\nabla_{\mu}n^\mu + \frac{2 \lambda L^2}{(D-3)(D-4)}\bigl(P-2\tilde{G}_{\mu\nu}\mathcal{K}^{\mu\nu} \bigr)
\label{surfaceGB}
\end{eqnarray}
where $\lambda$ is the coefficient of the Gauss-Bonnet term, $P$ is the trace of the following tensor
\begin{eqnarray}
P_{\mu\nu}=\frac{1}{3} \bigl(2\mathcal{K} \mathcal{K}_{\mu\rho}\mathcal{K}^{\rho}_{\ \nu} + \mathcal{K}_{\rho\sigma}\mathcal{K}^{\rho\sigma}\mathcal{K}_{\mu\nu} 
-2 \mathcal{K}_{\mu\rho}\mathcal{K}^{\rho\sigma}\mathcal{K}_{\sigma\nu}-\mathcal{K}^2 \mathcal{K}_{\mu\nu} \bigr)
\label{traceP}
\end{eqnarray}
and $\tilde{G}_{\mu\nu}$ stands for Einstein tensor of $d$-dimensional boundary metric. For a Gauss-Bonnet black hole, the $D$-dimensional metric is given by
\begin{eqnarray}
&& ds^2 = -f(r) N^2 dt^2+\frac{1}{f(r)}dr^2+r^2d\Sigma^{2}_{d-1} \ , \nonumber\\
&& f(r)=-1+\frac{r^2}{2\lambda L^2}\biggl[1-\sqrt{1-4\lambda + \frac{4\lambda m}{r^{D-1}}}   \biggr]
\label{GBmetric}
\end{eqnarray}
where $m$ is related to the mass of the black hole. Performing the analogous steps as in the previous subsection to calculate the charge and Virasoro
algebra for Gauss-Bonnet black hole, we get
\begin{eqnarray}
Q[\xi_m]=\frac{1}{8 \pi G_{D}} \biggl(1-\frac{2(D-2)\lambda L^2}{(D-4)r_{h}^2}  \biggr) \int_{\mathcal{H}} d^{d-1}x \sqrt{h} \bigg(\kappa T_m - \frac{1}{2}\partial_{t}T_m \biggr)
\label{chargeGB}
\end{eqnarray}
\begin{eqnarray}
\bigl[Q[\xi_m],Q[\xi_n]\bigr]=\frac{1}{8 \pi G_{D}}  \biggl(1-\frac{2(D-2)\lambda L^2}{(D-4)r_{h}^2}  \biggr) \int_{\mathcal{H}} d^{d-1}x \sqrt{h}\biggl[\kappa \bigg(T_m\partial_{t}T_n-T_n\partial_{t}T_m \biggr) \nonumber\\
-\frac{1}{2}\biggl(T_{m}\partial_{t}^{2} T_{n} - T_{n}\partial_{t}^{2}T_{m} \biggr) +\frac{1}{4\kappa}\biggl(\partial_{t}T_m\partial_{t}^{2}T_n-\partial_{t}T_n\partial_{t}^{2}T_m \biggr) \biggr]
\label{chargeBraGB}
\end{eqnarray}
Now substituting eqs.(\ref{Tm}) and (\ref{GBmetric}) into eqs.(\ref{chargeGB}) and (\ref{chargeBraGB}), we obtain
\begin{eqnarray}
&& Q[\xi_m]=\frac{A_{d-1}}{8 \pi G_{D}} \biggl(1-\frac{2(D-2)\lambda L^2}{(D-4)r_{h}^2}  \biggr) \frac{\kappa}{\alpha}\delta_{m,0} \nonumber\\
&& \bigl[Q[\xi_m],Q[\xi_n]\bigr]= -i(m-n)\frac{A_{d-1}}{8 \pi G_{D}}\frac{\kappa}{\alpha}\biggl(1-\frac{2(D-2)\lambda L^2}{(D-4)r_{h}^2}  \biggr)\delta_{m+n,0} \nonumber\\
&& \hspace{3cm} -im^3\frac{A_{d-1}}{16 \pi G_{D}}\frac{\alpha}{\kappa}\biggl(1-\frac{2(D-2)\lambda L^2}{(D-4)r_{h}^2}  \biggr)\delta_{m+n,0}
\label{Q0finalGB}
\end{eqnarray}
from which, we obtain the central term
\begin{eqnarray}
&& K\bigl[\xi_m,\xi_n\bigr]=\bigl[Q[\xi_m],Q[\xi_n]\bigr]+i(m-n)Q[\xi_{m+n}] \nonumber\\
&&  \hspace{2cm} = -im^3\frac{A_{d-1}}{16 \pi G_{D}}\frac{\alpha}{\kappa}\biggl(1-\frac{2(D-2)\lambda L^2}{(D-4)r_{h}^2}  \biggr)\delta_{m+n,0}
\label{centralKGB}
\end{eqnarray}
and we can read off the zero mode $Q_0$ and central charge $c$ as
\begin{eqnarray}
Q_0 = \frac{A_{d-1}}{8 \pi G_{D}}\frac{\kappa}{\alpha} \biggl(1-\frac{2(D-2)\lambda L^2}{(D-4)r_{h}^2}  \biggr), \ \
\frac{c}{12}= \frac{A_{d-1}}{16 \pi G_{D}}\frac{\alpha}{\kappa} \biggl(1-\frac{2(D-2)\lambda L^2}{(D-4)r_{h}^2}  \biggr)
\label{QoCGB}
\end{eqnarray}
Substituting above expressions into eq.(\ref{DOS2}) and taking a logarithm, we obtain the black hole entropy
\begin{eqnarray}
&& S=\ln{\rho}\approx 2\pi\sqrt{\frac{c\Delta}{6}} + \frac{1}{4}\ln{\frac{c}{\Delta^3}} +... \nonumber\\
&& S=\frac{A_{d-1}}{4 G_D}\biggl(1-\frac{2(D-2)\lambda L^2}{(D-4)r_{h}^2}  \biggr) -\frac{3}{2} \ln{\biggl[ \frac{A_{d-1}}{4G_D}\biggl(1-\frac{2(D-2)\lambda L^2}{(D-4)r_{h}^2}  \biggr)\biggr]}
+ ...\nonumber\\
&& = S_{Wald}-\frac{3}{2} \ln{S_{Wald}} +...
\label{GBentropy}
\end{eqnarray}
In the above equation, the first term is exactly the Wald entropy of a Gauss-Bonnet black hole in $D$-dimensions. We also find the logarithmic correction to the Wald entropy with 
coefficient $-3/2$, which is same as in the AdS-Schwarzschild black hole case. As one can see, however now in terms of horizon radius, there are two correction terms. 

Before ending this section, it is worthwhile to point out again the debatable nature of the coefficient of logarithmic term in black hole entropy. The logarithmic correction
to the usual black hole entropy has been discussed in various quantum gravity models. This correction term has been successfully computed for asymptotically flat
black holes in loop quantum gravity (LQG), in string theory for both extremal and near extremal black holes by microscopic counting methods or by diffeomorphism symmetry arguments which
heavily rely on two dimensional CFT and the Cardy formula (as we have done in this
section). All of these methods have either predicted different coefficient for the logarithmic term or have a certain level of arbitrariness in their results. 
For example, in LQG the coefficient was found to be $-3/2$ \cite{Kaul}. However
in string theory, this coefficient was found to be different from $-3/2$ \cite{Sen1005}-\cite{Bhattacharyya1210}. There, the logarithmic corrections to black hole entropy arise from the one
loop contribution of the massless fields, and the coefficient was found to be dependent on the total number of massless fields as well as the number of spacetime dimensions.

For the BTZ black hole, Carlip calculated this coefficient using the Cardy formula and found it to be $-3/2$ \cite{Carlip2000}. He also argued that similar logarithmic correction should 
appear in all black holes whose microscopic degrees
of freedom are described by an underlying horizon CFT. However in his calculation Carlip had assumed, as we have also assumed in this paper, that the central charge $c$ is universal constant
in a sense that it is independent of horizon area. To achieve this assumption Carlip chose the parameter $\alpha$ (appearing in eq.(\ref{Tm})) in such a way that $c$ becomes independent of area.
It is important to point out here that one could choose $\alpha$ to be the surface gravity $\kappa$ as well. The condition $\alpha=\kappa$ is also well motivated from Euclidean gravity point of view and
is standard in the literature \footnote{In various methods of computing black hole entropy from diffeomorphism symmetry arguments, the parameter $\alpha$ is generally chosen to be the 
surface gravity by hand to get the correct expression for black hole entropy, see for instance \cite{Carlip9812}\cite{Majhi1111}\cite{Silva0204}.}. To see this, let us consider
the Euclidean time $(\tau\rightarrow it)$ and take the appropriate ansatz for $T_m$ as $T_m=1/\alpha e^{i m (\alpha\tau + g(\tilde{r}) + \vec{p}.\vec{x})}$. In the Euclidean
formalism our analysis still go through, but now $\tau$ must have a periodicity of $2 \pi/\kappa$ to avoid the conical singularity. In order to maintain this periodicity in $\tau$, 
we must choose $\alpha=\kappa$. If we choose this condition then its not hard to see that the coefficient of logarithmic correction is $-1/2$ instead of $-3/2$.

From above discussion, it is therefore fair to say that as of now there is no consensus on the coefficient of this logarithmic correction term and that more work in needed in order to say 
anything conclusively. Since we are mostly interested in studying the effects of this logarithmic correction term on the holographic R\'{e}nyi entropy, we adopt here a more neutral point of view and 
consider a general logarithmic correction with arbitrary coefficient $\textbf{C}$, instead of worrying too much about its exact magnitude
\begin{eqnarray}
S = S_{Wald} + \textbf{C}  \ln{S_{Wald}}+...
\label{Waldentropymain}
\end{eqnarray}
The whole purpose of our previous exercise, where we computed $\textbf{C}=-3/2$, was to show at-least one method by which logarithmic correction to the entropy of topological black holes, not just
in Einstein gravity but in other higher derivative gravity too, can be explicitly calculated and that this coefficient is non-zero. It would certainly be interesting to generalize the 
method of \cite{Kaul} in LQG or of \cite{Sen1205} in string theory to calculate logarithmic correction for topological black holes. Especially, for higher derivative gravity theories it is not clear (at-least to
the author) how to proceed. It would be an interesting problem in its own right to analyze similarities and differences in the results of these methods but it is beyond the scope of this paper.

\section{Calculations and Results}
In this section, we will present results for the holographic R\'{e}nyi entropy by considering logarithmic correction to the usual black hole entropy. We will again concentrate on two gravity theories: Einstein and Gauss-Bonnet. Our main focus here will be to see the effects of this logarithmic correction on the R\'{e}nyi entropy. However, before going into the details of each case separately,
let us first calculate the area of the hyperbolic event horizon appearing in eqs.(\ref{SAdSentropy}) and (\ref{GBentropy}). The line element on the $(d-1)$-dimensional hyperbolic
space is given
in eq.(\ref{hyperboliclineelement}), which we reproduce here again for convenience
\begin{eqnarray}
d\Sigma^{2}_{d-1}= d\theta^2 + \sinh^2\theta \ d\varOmega_{d-2}^2
\end{eqnarray}
It would be convenient if we make a change of coordinate $\theta=\cosh^{-1}y$, in which case the above metric reduces to
\begin{eqnarray}
d\Sigma^{2}_{d-1}= \frac{dy^2}{y^2-1} + (y^2-1) d\varOmega_{d-2}^2
\end{eqnarray}
The area of the hyperbolic space (and hence the area of the event horizon) is divergent. In order to regulate this area, we introduce a upper cutoff by integrating out to a maximum 
radius
\begin{eqnarray}
y_{max}=\frac{R}{\delta}
\label{cutoff}
\end{eqnarray}
where $\delta$ is the short distance cutoff related to the UV cutoff of the boundary CFT. Now, the area $V_{\Sigma_{d-1}}$ of the hyperbolic space is calculated as
\begin{eqnarray}
&& V_{\Sigma_{d-1}}=\Omega_{d-2} \int_{1}^{y_{max}}dy \ (y^2-1)^{(d-3)/2} \nonumber\\
&&  \simeq \frac{\Omega_{d-2}}{d-2} \biggl[(y_{max}^{d-2}-1)-\frac{(d-2)(d-3)}{2(d-4)}(y_{max}^{d-4}-1) \nonumber\\ 
&& \hspace{6.7cm} +\frac{(d-2)(d-3)(d-5)}{8(d-6)}(y_{max}^{d-6}-1)-...  \biggr] \nonumber\\
&& \simeq \frac{\Omega_{d-2}}{d-2} \biggl[\frac{R^{d-2}}{\delta^{d-2}}-\frac{(d-2)(d-3)}{2(d-4)}\frac{R^{d-4}}{\delta^{d-4}}+\frac{(d-2)(d-3)(d-5)}{8(d-6)}\frac{R^{d-6}}{\delta^{d-6}}-...\biggr]
 \nonumber\\
\label{volumehyper}
\end{eqnarray}
where $\Omega_{d-2}=2 \pi^{(d-1)/2}/\varGamma((d-1)/2)$ is the area of a unit $(d-2)$ sphere. However, for $d=2$,  we have a logarithmic behaviour
\begin{eqnarray}
&& V_{\Sigma_{1}}= 2 \int_{1}^{y_{max}}dy \ (y^2-1)^{-1/2}=2\ln{(2y_{max})}=2\ln{\biggl(\frac{2R}{\delta}\biggr)}
\label{volumehyper1}
\end{eqnarray}
Therefore, the area of the event horizon $A_{d-1}$ is given by
\begin{eqnarray}
A_{d-1}= r_{h}^{d-1} V_{\Sigma_{d-1}}
\label{eventhorizonSAdS}
\end{eqnarray}
Now we have all the ingredients to compute the holographic R\'{e}nyi entropy for spherical entangling surface.

\subsection{R\'{e}nyi entropy from AdS-Schwarzschild Black hole}

For AdS-Schwarzschild Black hole the metric is given in eq.(\ref{AdSSch}). However, in order to make comparison with \cite{Hung1110}, let us rewrite this in a slightly different form
\begin{eqnarray}
&& ds^2 = -[-1+\frac{r^2}{L^2}g(r)]N^2 dt^2+\frac{1}{[-1+\frac{r^2}{L^2}g(r)]}dr^2 + r^2d\Sigma^{2}_{d-1} \ , \nonumber\\
&& \hspace{3cm} g(r)=\frac{L^2}{r^2}(f(r)+1)
\label{AdSSch1}
\end{eqnarray}
In order to ensure that the boundary spacetime is conformally equivalent to $R\times H^{d-1}$, \textit{i.e}
\begin{eqnarray}
ds_{\infty}^2 = -dt^2+R^2d\Sigma^{2}_{d-1}
\end{eqnarray}
we choose the constant $N^2=L^2/(g_{\infty}R^2)=\tilde{L}^2/R^2$, where $g_{\infty}=\lim_{r\to\infty}g=1$.
The Hawking temperature of this black hole is given by 
\begin{eqnarray}
T=\frac{1}{4 \pi R}\biggl(\frac{d r_h}{L} -\frac{(d-2)L}{r_h} \biggr)
\label{HTSAdS}
\end{eqnarray}
which is also the temperature of the boundary CFT on $R\times H^{d-1}$. For computational purposes, it is convenient to consider a coordinate $x=r_h/\tilde{L}$ in which case the expression
for the R\'{e}nyi entropy in eq.(\ref{Sqmain}) reduces to
\begin{eqnarray}
&& S_{q}= \frac{q}{q-1}\frac{1}{T_0} \int_{x_q}^{1} dx \  S_{therm}(x)\frac{dT}{dx} \nonumber\\
&& = \frac{q}{q-1}\frac{1}{T_0} \biggl[ S_{therm}(x)T(x)\rvert_{x_q}^{1} - \int_{x_q}^{1} dx \  T(x) \frac{d}{dx}S_{therm}     \biggr]
\label{Sqmain1}
\end{eqnarray}
Here, in the above integral, the upper limit $x=1$ comes from the condition $T=T_0$ which implies $r_h=\tilde{L}$. The lower limit $x_q$, which needs to be determined, corresponds to the
temperature $T=T_0/q$. Using eq.(\ref{HTSAdS}), $x_q$ must satisfy the following equation 
\begin{eqnarray}
d x_{q}^2 -\frac{2}{q} x_q - (d-2)=0
\end{eqnarray}
the real and positive root of which is given by
\begin{eqnarray}
x_{q}= \frac{1}{qd}\biggl[1+\sqrt{d^2 q^2 +1 -2dq^2}  \biggr]
\label{xqSAdS}
\end{eqnarray}
Finally, using the black hole entropy expression (eq.(\ref{Waldentropymain})) in place of $S_{therm}$ in eq.(\ref{Sqmain1}), we get the expression for R\'{e}nyi entropy as 
\begin{eqnarray}
&& S_{q}= \frac{q \mathfrak{B}}{2(q-1)}(2 - x_{q}^{d-2}(1+x_{q}^2)) +\frac{\textbf{C} q}{q-1} \ln{\mathfrak{B}} + \nonumber\\
&& \frac{\textbf{C} q}{2(q-1)x_q} \biggl[(d-1)(x_q-1)(dx_q+2-d) - (dx_{q}^{2}+2-d) \ln(\mathfrak{B} x_{q}^{d-1})  \biggr]
\label{RESAdS}
\end{eqnarray}
where $\mathfrak{B}=2 \pi V_{\Sigma_{d-1}}(\frac{\tilde{L}}{l_p})^{d-1}$, $l_p$ is the Plank length related to the gravitation constant $G_D=l_{p}^{d-1}/8 \pi$. We see that $S_q$ is a complicated
function of $x_q$. For $\textbf{C}=0$, $S_q$ reduces to the expression found in \cite{Hung1110}. However, for $\textbf{C}\neq0$ there are additional nontrivial correction terms. We now make some 
observations: \\

\noindent $\bullet$ For $d=2$, we get 
\begin{eqnarray}
&& S_{q}(d=2)= \frac{\mathfrak{B}}{2} \biggl(1+\frac{1}{q} \biggr) + \textbf{C}\biggl(\ln (\mathfrak{B}) -1\biggr) +\textbf{C}\frac{\ln q}{q-1} \nonumber\\
&& =\frac{c}{6}\biggl(1+\frac{1}{q} \biggr)\ln{\biggl(\frac{2R}{\delta}\biggr)}+ \textbf{C}\biggl[ \ln \biggl(\frac{c}{3}\ln \biggl(\frac{2R}{\delta}\biggr) \biggr) -1  \biggr]
+\textbf{C}\frac{\ln q}{q-1}
\label{RESAdSd2}
\end{eqnarray}
where $c=12\pi L/l_p$ is the standard expression of the central charge in the two dimensional boundary CFT \footnote{This $c$ should not be confused with the central charge of eq.(\ref{QoCSAdS})
which appears in the Virasoro algebra on the horizon. The $c$ in eq.(\ref{QoCSAdS})
was used to compute the black hole entropy and has nothing to do with the boundary CFT.}. The first term in eq.(\ref{RESAdSd2}) matches with the well known result 
of the R\'{e}nyi entropy in a two dimensional CFT for an interval of length $l=2R$. This is expected as, for $d=2$, the spherical entangling region consists of two points separated
by distance $2R$. However, our main result here is the appearance of additional corrections to R\'{e}nyi entropy which are precisely coming from the coefficient of the logarithmic correction 
to black hole entropy. These correction terms to the R\'{e}nyi entropy have both additional logarithmic as well as double logarithmic structure. \\

\noindent $\bullet$ Similarly, the entanglement entropy for $d=2$ is obtained by taking $q\rightarrow1$ limit of eq.(\ref{RESAdSd2})
\begin{eqnarray}
S_{1}(d=2) = \frac{c}{3}\ln{\biggl(\frac{2R}{\delta}\biggr)}+ \textbf{C}\biggl[ \ln \biggl(\frac{c}{3}\ln \biggl(\frac{2R}{\delta}\biggr) \biggr) \biggr]
\label{EESAdSd2}
\end{eqnarray} 
The leading term again has the same structure as is well known for the two dimensional CFT. However, now we have found  additional correction too. Interestingly, the correction term in the 
entanglement entropy has a simple structure. It is proportional to $\textbf{C}$ times the logarithm of its standard expression. Moreover, the same is true in all dimensions. That is,
\begin{eqnarray}
S_{1}(d) = \mathfrak{B} + \textbf{C}\ln[\mathfrak{B}] = 2 \pi V_{\Sigma_{d-1}}(\frac{\tilde{L}}{l_p})^{d-1} + \textbf{C} \ln[2 \pi V_{\Sigma_{d-1}}(\frac{\tilde{L}}{l_p})^{d-1}] \nonumber\\
=  2 \pi V_{\Sigma_{d-1}}(\frac{\tilde{L}}{l_p})^{d-1} + \textbf{C}  \ln[(\frac{\tilde{L}}{l_p})^{d-1}] + \textbf{C} \ln[2 \pi V_{\Sigma_{d-1}}]
\label{EESAdSd}
\end{eqnarray} 
In the last line, we have divided the correction term into two parts. One that depends on $(\tilde{L}/l_p)^{d-1}$, which is related to the number of degrees of freedom of the boundary CFT
and the other that depends on the size of the system. Let us concentrate on the term which is of order $G_{D}^0$ (last term in eq.(\ref{EESAdSd})) 
\footnote{Quantum corrections to the RT formula are expected to occur at the order $G_{D}^0$ \cite{Faulkner1307}. The RT formula gives entanglement entropy at the order $1/G_D$.
See the discussion section for more details on this.}.
We observe that at this order there is a logarithmic correction in all dimensions. In particular, there is a logarithmic correction in odd $d$ as well.
In order to see this, we write the complete expression for entanglement entropy in $d=3$ and $4$ dimensions
\begin{eqnarray}
S_{1}(d=3) = \biggl(2 \pi \frac{\tilde{L}}{l_p}\biggr)^2 \biggl(\frac{R}{\delta}-1 \biggr) + \textbf{C} \ln \biggl(\frac{\tilde{L}}{l_p} \biggr)^2 
+ \textbf{C} \ln \biggl(4 \pi^2 \frac{R}{\delta} \biggr) +...   \nonumber\\
S_{1}(d=4) = 4 \pi^2 \biggl(\frac{\tilde{L}}{l_p}\biggr)^3 \biggl(\frac{R^2}{\delta^2}-\ln\frac{2R}{\delta} \biggr) + \textbf{C} \ln \biggl(\frac{\tilde{L}}{l_p} \biggr)^3 
+  \textbf{C} \ln \biggl(4 \pi^2 \frac{R^2}{\delta^2} \biggr) +...
\label{EESAdSd3d4}
\end{eqnarray} 
we note from the last terms of eq.(\ref{EESAdSd3d4}) that at the order $G_{D}^0$ there is a correction to the entanglement entropy which depends logarithmically on the size of the entangling
surface. The same result holds in all $d>2$ dimensions.
\\

\begin{figure}[t!]
\begin{minipage}[b]{0.5\linewidth}
\centering
\includegraphics[width=2.8in,height=2.3in]{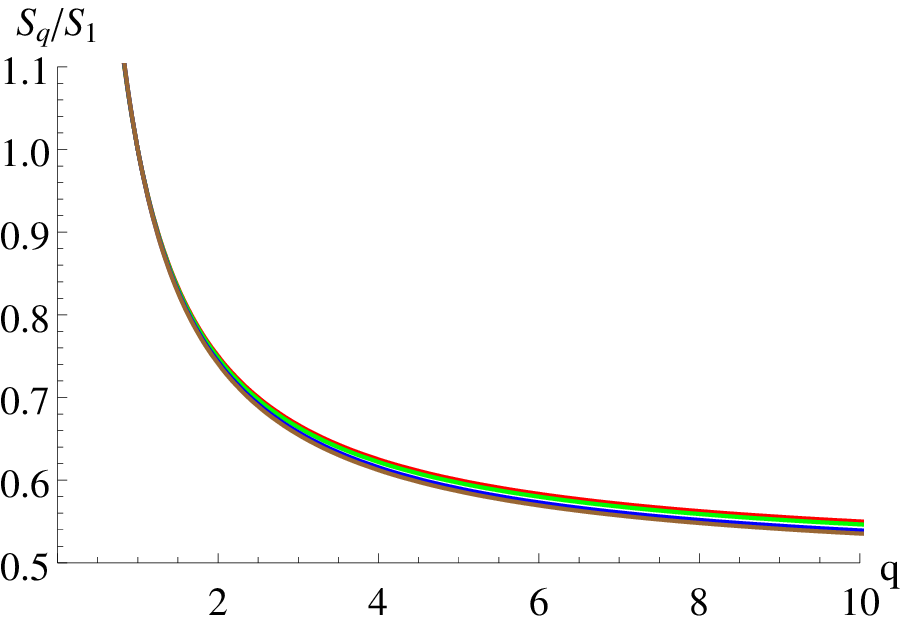}
\caption{ \small $S_q/S_1$ as a function of $q$ for various $\textbf{C}$. Here $d=2$ and red, green, blue and brown curves correspond to 
$\textbf{C}$=$0$, $-1/2$, $-3/2$ and $-2$ respectively.}
\label{SqbyS1Vsqd2withCSAdS}
\end{minipage}
\hspace{0.4cm}
\begin{minipage}[b]{0.5\linewidth}
\centering
\includegraphics[width=2.8in,height=2.3in]{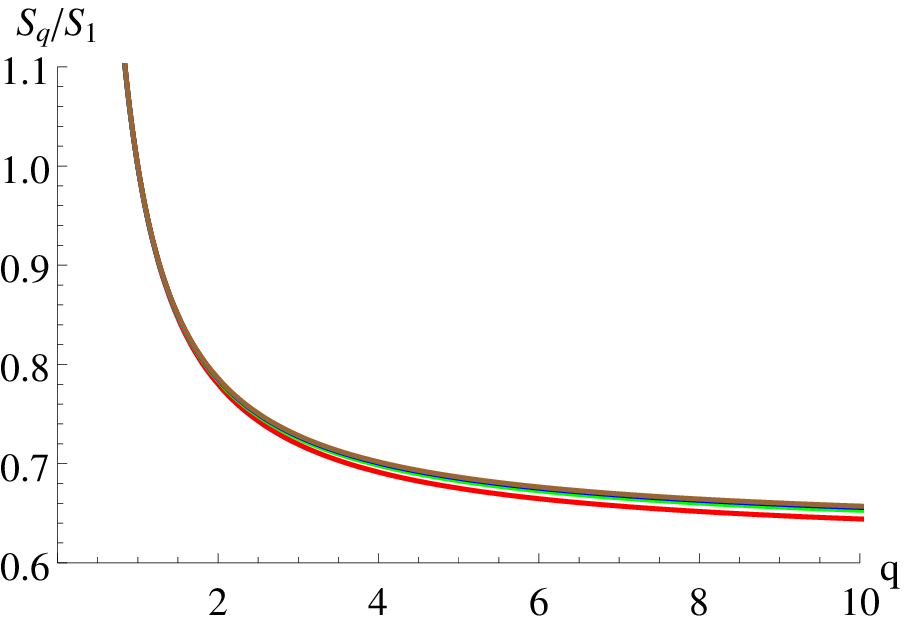}
\caption{\small $S_q/S_1$ as a function of $q$ for various $d$. Here $\textbf{C}=-3/2$ and red, green, blue and brown curves correspond to 
$d$= $3$, $4$, $5$ and $6$ respectively.}
\label{SqbyS1VsqVsdwithCN1by2SAdS}
\end{minipage}
\end{figure}

\noindent $\bullet$ For $d=2$, the ratio of R\'{e}nyi entropy to entanglement entropy as a function of $q$ for some reasonable values of $\textbf{C}$ is plotted in fig.(\ref{SqbyS1Vsqd2withCSAdS}). 
In order to plot this ratio, we have chosen $R=1$, $\delta=10^{-4}$ and $L=2l_p$. We see that the overall behaviour of $S_q/S_1$ is same for all $\textbf{C}$. For a fixed $q>1$, 
the magnitude of the ratio 
$S_q/S_1$ decreases with decrease in $\textbf{C}$. However, we should emphasize here that these results are cutoff ($\delta$) dependent. For instance, for much smaller cutoff,
say $\delta=10^{-20}$, differences due to
$\textbf{C}\neq 0$ are almost negligible. Similarly, we also found that for larger and larger values of $c$ (or $L/l_p$), the effect of $\textbf{C}$ become smaller and smaller. We can also 
note that $S_q/S_1>1$ for $q<1$ and $S_q/S_1<1$ for $q>1$. \\

\noindent $\bullet$ Let us also note some useful limits of the R\'{e}nyi entropy
\begin{eqnarray}
&& S_{\infty}=\mathfrak{B} \biggl[1-\frac{d-1}{d} \biggl(\frac{d-2}{d}\biggr)^{\frac{d-2}{2}} \biggr] + \textbf{C}\biggl[\ln \mathfrak{B} -(d-1)^2 + (d-1)\sqrt{d(d-2)} \biggr] \nonumber\\
&& S_{1}= \mathfrak{B} + \textbf{C} \ln \mathfrak{B}  \nonumber\\
&& S_{0} = \mathfrak{B} \biggl[ \biggl(\frac{2}{d} \biggr)^{d} \frac{1}{2 q^{d-1}} + \biggl(\frac{2}{d} \biggr)^{d-2} \frac{1}{2 q^{d-3}}   \biggr] 
+ \textbf{C}\biggl[\ln \mathfrak{B} -(d-1) - (d-1)\ln (dq)     \biggr] \nonumber\\
\label{EESAdSvariouslimits}
\end{eqnarray}
\\
\noindent $\bullet$ Now, we discuss R\'{e}nyi and entanglement entropies for higher dimensional theories. In fig.(\ref{SqbyS1VsqVsdwithCN1by2SAdS}), $S_q/S_1$ as a 
function of $q$ for various values of $d$ is shown. Here, we have considered $\textbf{C}=-3/2$ and again 
have chosen $R=1$, $\delta=10^{-4}$ and $L=2l_p$. The overall characteristic features of $S_q/S_1$ is found to be same as in the $d=2$ case.
We found that, for fixed $\textbf{C}$, as we increase the number of spacetime dimensions the magnitude of the ratio $S_q/S_1$ increases. Even for large $d$, we again found 
that $S_q/S_1>1$ for $q<1$ and $S_q/S_1<1$ for $q>1$. \\

\noindent $\bullet$ Let us also comment on some of the useful inequalities of the R\'{e}nyi entropy. It is well known that R\'{e}nyi entropy satisfies inequalities involving the derivative 
with respect $q$, such as
\begin{eqnarray}
&& \partial_{q}S_{q}\leq 0 \nonumber\\
&& \partial_{q}[\frac{q-1}{q}S_q]\geq 0 \nonumber\\
&& \partial_{q}[(q-1)S_q]\geq 0 \nonumber\\
&& \partial_{q}^2 [(q-1)S_q]\leq 0
\label{Renyiinequal}
\end{eqnarray}
therefore, it is natural to ask whether or not these inequalities are satisfied with the corrected results. For $d>2$, we find that these inequalities are indeed 
satisfied provided $\textbf{C}$ is not very large (which is expected for small corrections). However for $d=2$, we find few subtleties, especially in the fourth inequality of
eq.(\ref{Renyiinequal}). We find that for small central charge it can be violated. Although for large central charge, which is the case for boundary theories that have a gravity dual,
the fourth inequality is again found to be satisfied.     \\

\noindent $\bullet$ Perhaps, the most important result form our computation is that the leading UV divergences appearing in eq.(\ref{RESAdS}) is same for all $q$. Indeed, if we note that
leading term $R^{d-2}\Omega_{d-2}$ in eq.(\ref{volumehyper}) gives the area of the spherical entangling region, then the leading divergent term always has a ``area law'' structure for $d>2$. 
For $d=2$, we have the usual logarithmic divergence in the R\'{e}nyi entropy. As expected, the correction terms do not change the leading behavior of the R\'{e}nyi entropy. \\

\subsection{R\'{e}nyi entropy from Gauss-Bonnet Black hole}
In this subsection, we examine the holographic R\'{e}nyi entropy for boundary theory which are dual to Gauss-Bonnet gravity theory. The procedure for calculating R\'{e}nyi entropy with 
Gauss-Bonnet Black hole background is entirely similar to what has been discussed in the previous subsection, and therefore we will be brief here.

The metric for Gauss-Bonnet Black hole is given in eq.(\ref{GBmetric}), however we rewrite it in a slightly different form in order to make comparison with \cite{Hung1110}
\begin{eqnarray}
&& ds^2 = -[-1+\frac{r^2}{L^2}g(r)]N^2 dt^2+\frac{1}{[-1+\frac{r^2}{L^2}g(r)]}dr^2 + r^2d\Sigma^{2}_{d-1} \ , \nonumber\\
&& g(r)= \frac{1}{2\lambda}\biggl[1-\sqrt{1-4\lambda + \frac{4\lambda m}{r^{d}}} \biggr]
\label{GBmetric1}
\end{eqnarray}
In this case $N^2=L^2/(g_{\infty}R^2)=\tilde{L}^2/R^2$. Since now $g_{\infty}=(1-\sqrt{1-4\lambda})/(2\lambda)\neq 1$, it implies $\tilde{L}\neq L$. This is just the statement that the 
AdS curvature scale is distinct from the length scale $L$ which usually appears in the Gauss-Bonnet gravity action. Now, again using the coordinate $x=r_h/\tilde{L}$,
the Hawking temperature can be expressed as
\begin{eqnarray}
T=\frac{1}{2\pi R x}\biggl(1+ \frac{d}{2 g_\infty} \frac{x^4 -x^2 g_\infty + \lambda g_{\infty}^2}{x^2- 2 \lambda g_\infty} \biggr)
\label{HTGB}
\end{eqnarray}
Similarly, the horizon entropy (eq.(\ref{GBentropy})) can be recast as
\begin{eqnarray}
S=2 \pi \biggl(\frac{\tilde{L}}{l_p}\biggr)^{d-1} V_{\Sigma_{d-1}} x^{d-1} \biggl(1-\frac{2\lambda g_\infty}{x^2} \frac{d-1}{d-3} \biggr) + 
\textbf{C} \ln \bigg[ 2 \pi \biggl(\frac{\tilde{L}}{l_p}\biggr)^{d-1} V_{\Sigma_{d-1}} x^{d-1} \times \nonumber\\
\biggl(1-\frac{2\lambda g_\infty}{x^2} \frac{d-1}{d-3} \biggr)  \biggr]
\label{entropyGB1}
\end{eqnarray}
After substituting above results into eq.(\ref{Sqmain1}), we can get expression for the R\'{e}nyi entropy for any $d$. However, for general $d$ the expressions for R\'{e}nyi entropy
is very complicated and lengthy and also not very illuminating. For this reason we focus on $d=4$ case, in which the R\'{e}nyi entropy has the following expression
\begin{eqnarray}
S_{q}= \frac{q \mathfrak{B}}{2(q-1)} \biggl[\frac{1-x_{q}^{3}}{g_\infty} -3 (1-x_{q}^{2}) + 4(1-4\lambda)\biggl(\frac{1}{1-2\lambda g_\infty} - \frac{x_{q}^4}{x_{q}^2-2\lambda g_\infty}\biggr)\biggr] 
+ \nonumber\\ 
\frac{\textbf{C}q (g_\infty -2)}{(q-1)g_\infty(2\lambda g_\infty -1)} \ln [\mathfrak{B} (1-6\lambda g_\infty)]-\frac{\textbf{C}q x_q(g_\infty -2 x_{q}^2)}{(q-1)g_\infty(2\lambda g_\infty -x_{q}^2)} 
 \ln [\mathfrak{B} x_{q}(x_{q}^2-6\lambda g_\infty)] \nonumber\\ +
\frac{6\textbf{C}q}{ (q-1)}\frac{(x_q-1)}{g_\infty} + \frac{\sqrt{3}\textbf{C}q (12\lambda-1)}{(q-1) \sqrt{2\lambda g_\infty}} \biggl[\tanh^{-1}\biggl(\frac{1}{\sqrt{6\lambda g_\infty}}  \biggr) 
-\tanh^{-1}\biggl(\frac{x_q}{\sqrt{6\lambda g_\infty}}  \biggr) \biggr]
\label{REGB}
\end{eqnarray}
where $x_q$ corresponds to the positive and real root of the following equation
\begin{eqnarray}
4 q x_{q}^4 -2g_{\infty}x_{q}^3 -2 q g_\infty x_{q}^2 + 4\lambda g_{\infty}^2 x_{q}=0
\end{eqnarray}
which is again obtained from the equation $T=T_0/q$, \textit{i.e.} for the lower limit of the integral in eq.(\ref{Sqmain1}). The $\lambda$ which appears in the argument of logarithmic 
and inverse hyperbolic functions in eq.(\ref{REGB}) should be understood as an absolute value.
The first term in eq.(\ref{REGB}) have the same expression for
R\'{e}nyi entropy as was found in \cite{Hung1110}. However now we also have additional correction terms, which are both logarithmic as well as non-logarithmic in nature. One can also
explicitly check that, in the limit $\lambda\rightarrow 0$, eq.(\ref{REGB}) reduces to eq.(\ref{RESAdS}) for Einstein gravity.
\\
\\
We now make some observations:

\noindent $\bullet$ It is well known that the dual four dimensional boundary CFT of five dimensional Gauss-Bonnet gravity theory has two distinct central charges \footnote{Again, 
these central charges should not be confused with  the central charge of eq.(43).}
\begin{eqnarray}
c=\pi^2\biggl(\frac{\tilde{L}}{l_p} \biggr)^3 (1-2\lambda g_\infty), \ \ \ \ a=\pi^2\biggl(\frac{\tilde{L}}{l_p} \biggr)^3 (1-6\lambda g_\infty)
\end{eqnarray}
In terms of these central charges the R\'{e}nyi entropy in eq.(\ref{REGB}) reduces to
\begin{eqnarray}
S_{q}= \frac{V_{\Sigma_{3}}}{4\pi}\frac{q (1-x_{q}^2)}{q-1} \biggl[ (5c-a)x_{q}^2 -(13c-5a) + 16c\frac{2cx_{q}^2-(c-a)}{(3c-a)x_{q}^2-(c-a)} \biggr]
\nonumber\\
+\frac{\textbf{C}q}{(q-1)}\ln\biggl[\frac{2 a V_{\Sigma_{3}}}{\pi}\biggr] - \frac{\textbf{C}q x_q}{(q-1)} \frac{((5c-a)x_{q}^2-3c+a)}{((3c-a)x_{q}^2-c+a)}
\ln\biggl[\frac{V_{\Sigma_{3}}x_q}{\pi} \biggl((3c-a)x_{q}^2-3c+3a\biggr) \biggr]  \nonumber\\
+\frac{2\textbf{C}q}{(q-1)}\frac{(3c^2+a^2-6ac)}{(3c-a)^2}\sqrt{\frac{3(3c-a)}{c-a}}\biggl[ \tanh^{-1}\sqrt{\frac{3c-a}{3(c-a)}}
-\tanh^{-1}\sqrt{\frac{3c-a}{3(c-a)}x_{q}^2} \biggr]  \nonumber\\ 
+ \frac{\textbf{C}q(x_q-1)}{(q-1)}\frac{(15c-3a)}{(3c-a)}
\label{REGB1}
\end{eqnarray}
which shows that the R\'{e}nyi entropy is quite a complicated function of these central charges. It is also clear that the
R\'{e}nyi entropy is not determined solely by the anomaly coefficient $a$ as in the case of entanglement entropy (see below). As in the case of Einstein gravity, here too, the size of the
entangling surface always appears logarithmic in the correction terms. This feature therefore seems to be a universal in nature. \\

\noindent $\bullet$ We get the expression for the entanglement entropy as,
\begin{eqnarray}
S_{1} = a \frac{2 V_{  \Sigma_{3}}}{\pi} + \textbf{C} \ln \biggl[ a \frac{2 V_{  \Sigma_{3}}}{\pi}  \biggr]
\end{eqnarray}
we see that, as in the Einstein gravity case, correction term to the entanglement entropy is still given by the logarithmic of its original expression. This result can be traced back to eq.(\ref{Sqmain}), where it is clear that entanglement entropy is nothing but the black hole entropy at temperature $T=T_0$.  Therefore, logarithmic correction to the black hole entropy implies logarithmic correction to the entanglement entropy. It is also easy to see that similar results hold in higher dimensions too. \\
\begin{figure}[t!]
\begin{minipage}[b]{0.5\linewidth}
\centering
\includegraphics[width=2.8in,height=2.3in]{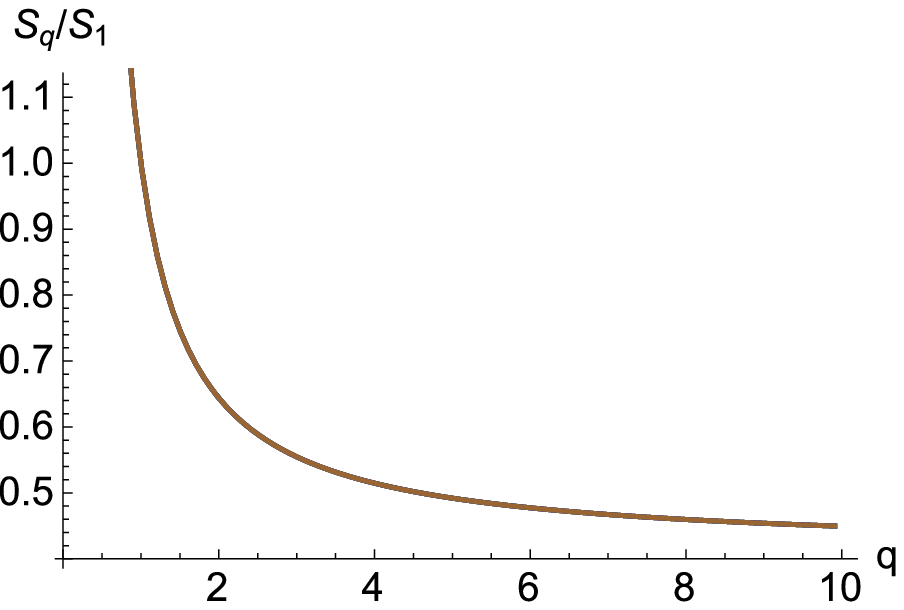}
\caption{ \small $S_q/S_1$ as a function of $q$ for various $\textbf{C}$. Here $\lambda=0.08$ and red, green, blue and brown curves correspond to 
$\textbf{C}$=$0$, $-1/2$, $-3/2$ and $-2$ respectively. For this plot, we have chosen $R=1$, $\delta=10^{-4}$ and $L=2l_p$.}
\label{SqbyS1LambdaPt08withalpha}
\end{minipage}
\hspace{0.4cm}
\begin{minipage}[b]{0.5\linewidth}
\centering
\includegraphics[width=2.8in,height=2.3in]{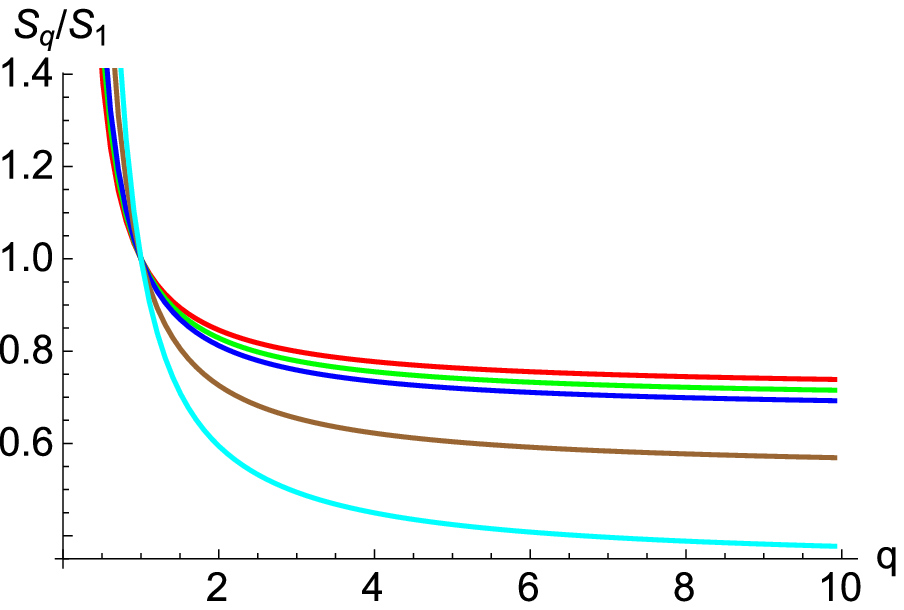}
\caption{\small $S_q/S_1$ as a function of $q$ for various $\lambda$. Here $\textbf{C}=-3/2$ and red, green, blue, brown and cyan curves correspond to 
$\lambda$= $-0.19$, $-0.1$, $-0.05$, $0.05$ and $0.09$ respectively. For this plot, we have chosen $R=1$, $\delta=10^{-4}$ and $L=2l_p$.}
\label{SqbyS1alphaN3by2withLambda}
\end{minipage}
\end{figure}

\noindent $\bullet$ $S_q/S_1$ as a function of $q$ for some reasonable values of $\textbf{C}$ and for $\lambda=0.08$ is shown in fig.(\ref{SqbyS1LambdaPt08withalpha}). We see that difference due to $\textbf{C}$ is extremely small and that the overall
behaviour of $S_q/S_1$ is same for all $\textbf{C}$. Similarly, $S_q/S_1$ for different values of $\lambda$ and for fixed $\textbf{C}=-3/2$ is shown in fig.(\ref{SqbyS1alphaN3by2withLambda}). We find that $S_q/S_1$, in the region $q<1$, increases for higher and higher values of $\lambda$, however in the region $q>1$, it decreases. Similar results holds for other values of $\textbf{C}$ as well. \\

\noindent $\bullet$ Before ending this section, let us also note some other useful limits of the R\'{e}nyi entropy,
\begin{eqnarray}
S_{0} = \frac{V_{\Sigma_{3}}}{4\pi}\frac{(3c-a)^4}{(5c-a)^3} \biggl(\frac{1}{q^3}+\frac{1}{q^2}\biggr)+\frac{V_{\Sigma_{3}}}{4\pi q}\frac{(3c-a)(5a^3-57a^2c+159ac^2-3c^3)}{(5c-a)^3} \nonumber\\
-3 \textbf{C} \ln q + \mathfrak{h}(c,a,V_{\Sigma_{3}})
\end{eqnarray}
\begin{eqnarray}
S_{\infty} = \frac{V_{\Sigma_{3}}}{\pi}\frac{(10ac-3c^2-2a^2)}{(5c-a)} + \textbf{C} \ln \biggl(2a\frac{V_{\Sigma_{3}}}{\pi}\biggr) +\frac{3\textbf{C}(5c-a)}{(3c-a)}
\biggl(\sqrt{\frac{3c-a}{5c-a}} -1\biggr) + \nonumber\\
\frac{2\textbf{C}(3c^2+a^2-6ac)}{(3c-a)^2}\sqrt{\frac{3(3c-a)}{c-a}}\biggl[\tanh^{-1}\sqrt{\frac{(3c-a)}{3(c-a)}} - \coth^{-1}\sqrt{3-\frac{12c^2}{(3c-a)^2}}  \biggr] 
\nonumber\\
\end{eqnarray}
where $\mathfrak{h}$ is some function of $c$, $a$ and $V_{\Sigma_{3}}$ but independent of $q$. The leading diverging term of $S_0$ matches exactly with \cite{Hung1110}.

\section{Discussions and Conclusions}

In this paper, we have studied the effects of logarithmic correction of the black hole entropy on the R\'{e}nyi entropy of a spherical entangling surface. We first used the diffeomorphism 
symmetry argument at the horizon to compute the black hole entropy expression. 
We used the Noether currents associated with the diffeomorphism invariance of the Gibbons-Hawking boundary action to construct the Virasoro algebra at the hyperbolic event horizons
and then used this algebra to calculate the entropy of AdS-Schwarzschild and Gauss-Bonnet black holes. We found that the leading term in the black hole entropy expression is the 
usual Wald entropy and that there is correction to it. This correction was found to be proportional to the logarithm of horizon area. 

We then applied the prescription of \cite{Casini1102}\cite{Hung1110} to calculate the holographic R\'{e}nyi entropy for a spherical entangling surface. Using the corrected black hole entropy 
expression we found that there are corrections to the standard expression of R\'{e}nyi entropy. These correction are shown in eq.(\ref{RESAdS}) for Einstein and in 
eq.(\ref{REGB1}) for Gauss-Bonnet gravity duals. In particular, we found that the R\'{e}nyi entropy is a complicated function of the index $q$ as well as the central charges. Interestingly, the
size of the entangling surface always appears logarithmically in the correction terms of the R\'{e}nyi entropy. This is true for both Einstein as well as Gauss-Bonnet gravity. We found that
the inequalities of R\'{e}nyi entropy are also satisfied even with correction terms.

It is important to analyze the nature and significance of these correction terms in the R\'{e}nyi entropy. If the corrections in the R\'{e}nyi entropy originating from the corrections in 
black hole entropy are quantum corrections, 
then our results can be useful in many directions. Especially since a lot of work have recently been appeared in the literature to compute leading quantum corrections
(of order $G_{D}^0$ or $N^0$) to the entanglement entropy holographically. In this context, 
a proposal for quantum corrections to holographic entanglement entropy is given in \cite{Faulkner1307}, also see \cite{Engelhardt1408}. In this proposal, the quantum corrections are
essentially given by the bulk entanglement entropy between RT minimal area surface and the rest of the bulk ( see fig.(1) of \cite{Faulkner1307}). As an example, the quantum correction to the 
entanglement entropy of the Klebanov-Strassler model in the large $N$ limit was calculated. The correction was found to be, as in our case, logarithmic in nature and it depends on the size of
the entangling surface. Indeed, one can also notice from
eq.(\ref{EESAdSd}) that at the order $G_{D}^0$ (recall that the factor $(\tilde{L}/l_p)^{d-1}$ measures the number of degrees of freedom of the dual CFT and it is related to the central
charge of the boundary CFT or to the $G_D$ on the gravity side) the correction to the entanglement entropy is proportional to the logarithm of the size of the entangling surface.
If the above interpretation is correct, then our results can be useful since they predict a similar kind of logarithmic correction to the R\'{e}nyi entropy.

An indirect hint for the logarithmic correction in the R\'{e}nyi entropy can also be seen in the following way. 
In \cite{Bhattacharyya1210}, a non-trivial test of the gauge/gravity 
duality at next-to-leading order in the $1/N$ expansion for ABJM theories was performed. There, it was shown that
the subleading logarithmic correction in the partition function (more correctly in $\log{Z}$) of the ABJM theory on a three sphere matches exactly with the partition function
of its eleven dimensional supergravity dual. The latter partition function at one-loop level was calculated using the Euclidean quantum gravity method.
The subleading logarithmic term in the partition function can lead to a logarithmic correction in the R\'{e}nyi
entropy, provided that, an analogous logarithmic term in the partition function on $q$-folded cover does not cancel in the 
definition of the R\'{e}nyi entropy. This scenario is partially true, at-least in odd dimensions.

Indeed one can notice, an important result of our analysis is that there are logarithmic corrections to entanglement and  R\'{e}nyi entropies in odd dimensions too. At first sight this 
seems strange as a logarithmic term in these entropies generally appears in even dimensions. However, it is also well known that the entanglement entropy for the sphere in flat 
space in odd dimensions is simply the negative of free energy on sphere \textit{i.e} $S_1=\log{Z}$ \cite{Klebanov1111}\cite{Dowker}, which implies that logarithmic corrections 
in the free energy directly lead to logarithmic
corrections in the entanglement entropy. As we have mentioned above, there are examples where logarithmic correction to the free energy indeed occur, therefore, it is 
not surprising that we got logarithmic correction in the entanglement entropy in odd dimensions as well. 

Since a similar kind of logarithmic subleading term in the partition function also arises in other boundary field theories, see for example \cite{Marino}, therefore it appears that
logarithmic correction to the R\'{e}nyi entropy might be a general feature of CFTs with gravity duals.
Although, in  order to explicitly establish this fact it would be useful if we can calculate the R\'{e}nyi entropy at one loop level on the lines of \cite{Barrella1306}
\footnote{In \cite{Barrella1306}, one-loop bulk corrections to RT formula was systematically calculated. This was done by calculating the one loop determinants around 
the classical solutions using the Schottky uniformization 
of $q$-sheeted Riemann surface. However there are many difficulties in implementing this method, such as constructing the smooth bulk solutions and performing analytic continuation 
of the replica index to non-integer $q$.}. 
It would certainly be interesting to explicitly compare the results of our calculations with those of \cite{Faulkner1307}\cite{Barrella1306}, and find the similarities and differences between 
them.

Finally, regardless of the problems associated with the interpretation of correction terms calculated in this paper, it is important to carefully examine and explore
their physical significance. Especially since in the context of AdS/CFT, it is well known that on the gravity side the logarithmic correction to black hole entropy 
arises only due to one loop contribution of the massless fields. Now, the mapping from the R\'{e}nyi entropy to the black hole entropy is exact and is expected to be valid at all orders.
Correspondingly, the logarithmic correction to black hole entropy must have some meaning in the R\'{e}nyi entropy too. Here, we have taken a small step in this direction and in the process have obtained several new results. We believe this mapping can be further useful, not just to better understand the structure 
of holographic R\'{e}nyi entropy but also to get a better understanding of the coefficient of the logarithmic correction in the gravity side. We hope to comment on this issue soon.

\begin{center}
{\bf Acknowledgements}
\end{center}
I am very grateful to B. Sathiapalan, N. Suryanarayana, R. Kaul and D. Dudal for useful discussions and for giving me valuable comments. I would like to thank
A. Dey, Zodinmawia, A. Sharma and D. Dudal for careful reading of the manuscript and pointing out the necessary corrections. Parts of the this work were done while the author was in The Institute of Mahematical Sciences, Chennai and was funded by a Postdoctoral fellowship provided by the Department of Atomic Energy of the Government of India. This work is partially supported by the postdoctoral grant PDM/15/172 from KU Leuven.

\end{document}